\documentclass[journal,twoside,web]{ieeecolor}
\usepackage{tmi}
\usepackage{cite}
\usepackage{amsmath,amssymb,amsfonts}
\usepackage{bbm}
\usepackage{algorithmic}
\usepackage{graphicx}
\usepackage{textcomp}
\usepackage{dsfont}
\usepackage[dvipsnames]{xcolor}

\usepackage{soul}
\usepackage{multirow}
\usepackage{booktabs}
\newcommand{\ra}[1]{\renewcommand{\arraystretch}{#1}}

\usepackage{caption}
\usepackage{hyperref}
\usepackage{float}
\graphicspath{ {./imgs/} }

\usepackage{xcolor}
\hypersetup{
    colorlinks,
    linkcolor={red!100},
    citecolor={blue!50!black},
    urlcolor={blue!80!black}
}

\def\BibTeX{{\rm B\kern-.05em{\sc i\kern-.025em b}\kern-.08em
    T\kern-.1667em\lower.7ex\hbox{E}\kern-.125emX}}
\begin{document}
\title{OCTAve: 2D \textit{en face} Optical Coherence Tomography Angiography Vessel Segmentation in Weakly-Supervised Learning with Locality Augmentation}
\author{Amrest Chinkamol \IEEEmembership{Member, IEEE}, Vetit Kanjaras, Phattarapong Sawangjai, Yitian Zhao, Thapanun Sudhawiyangkul, Chantana Chantrapornchai$^{*}$, Cuntai Guan \IEEEmembership{Fellow, IEEE} and Theerawit Wilaiprasitporn$^{*}$ \IEEEmembership{Member, IEEE}

\thanks{This work was supported by PTT Public Company Limited, The SCB Public Company Limited, The Office of the Permanent Secretary of the Ministry of Higher Education, Science, Research and Innovation, Thailand (RGNS63-252) and National Research Council of Thailand (N41A640131)  \textit{($^{*}$corresponding author: Chantana Chantrapornchai, Theerawit Wilaiprasitporn).}}
\thanks{A. Chinkamol, P. Sawangjai, T. Sudhawiyangkul and T. Wilaiprasitporn are with Bio-inspired Robotics and Neural Engineering (BRAIN) Lab, School of Information Science and Technology (IST), Vidyasirimedhi Institute of Science \& Technology (VISTEC), Rayong, Thailand (e-mail: amrest.c@vistec.ac.th, theerawit.w@vistec.ac.th).}
\thanks{V. Kanjaras is with Kamnoetvidya Science Academy (KVIS), Rayong, Thailand (e-mail: vetit\_k@kvis.ac.th).}
\thanks{Y. Zhao is with the Cixi Institute of Biomedical Engineering, Ningbo Institute of Industrial Technology, Chinese Academy of Sciences, Ningbo 315201, China}
\thanks{C. Chantrapornchai is with High-Performance Computing and Networking Center (HPCNC), Department of Computer Engineering, Faculty of Engineering, Kasetsart University, Bangkhen Campus, Bangkok, Thailand (email: fengcnc@ku.th)}
\thanks{C. Guan is with the School of Computer Science and Engineering, Nanyang Technological University, Singapore}}
\maketitle

\begin{abstract}
\textcolor{black}{While} there have been increased researches using deep learning techniques for the extraction of vascular structure from the 2D \textit{en face} OCTA, \textcolor{black}{for such approach, it is known that   
the data annotation process} \textcolor{black}{on the curvilinear structure like the retinal vasculature} \textcolor{black}{is very costly and time consuming}\textcolor{black}{, albeit few tried to address the annotation problem.}
 \textcolor{black}{In this work}, we \textcolor{black}{propose  the application of the scribble-base weakly-supervised learning method to automate the}
 pixel-level annotation.  The proposed method, called OCTAve, \textcolor{black}{combines  the weakly-supervised learning  using scribble-annotated ground truth} augmented with an adversarial and a novel self-supervised deep supervision.    Our \textcolor{black}{novel} \textcolor{black}{mechanism is designed to utilize the discriminative outputs from the discrimination layer of a UNet-like architecture} where the  Kullback-Liebler Divergence \textcolor{black}{between the aggregate discriminative outputs and the segmentation map predicate is minimized during the training. This combined method} leads to the better localization of the vascular structure as shown in our experiments. We  validate our proposed method on the large public datasets \textcolor{black}{i.e., ROSE, OCTA-500}. The segmentation performance is compared against  \textcolor{black}{both} state-of-the-art fully-supervised and \textcolor{black}{scribble-based weakly-supervised}  approaches. The implementation of our work used in the experiments is located at [LINK].
\end{abstract}


\section{Introduction}
\label{sec:introduction}


\begin{figure}[]
    \centering
    \includegraphics[width=\columnwidth]{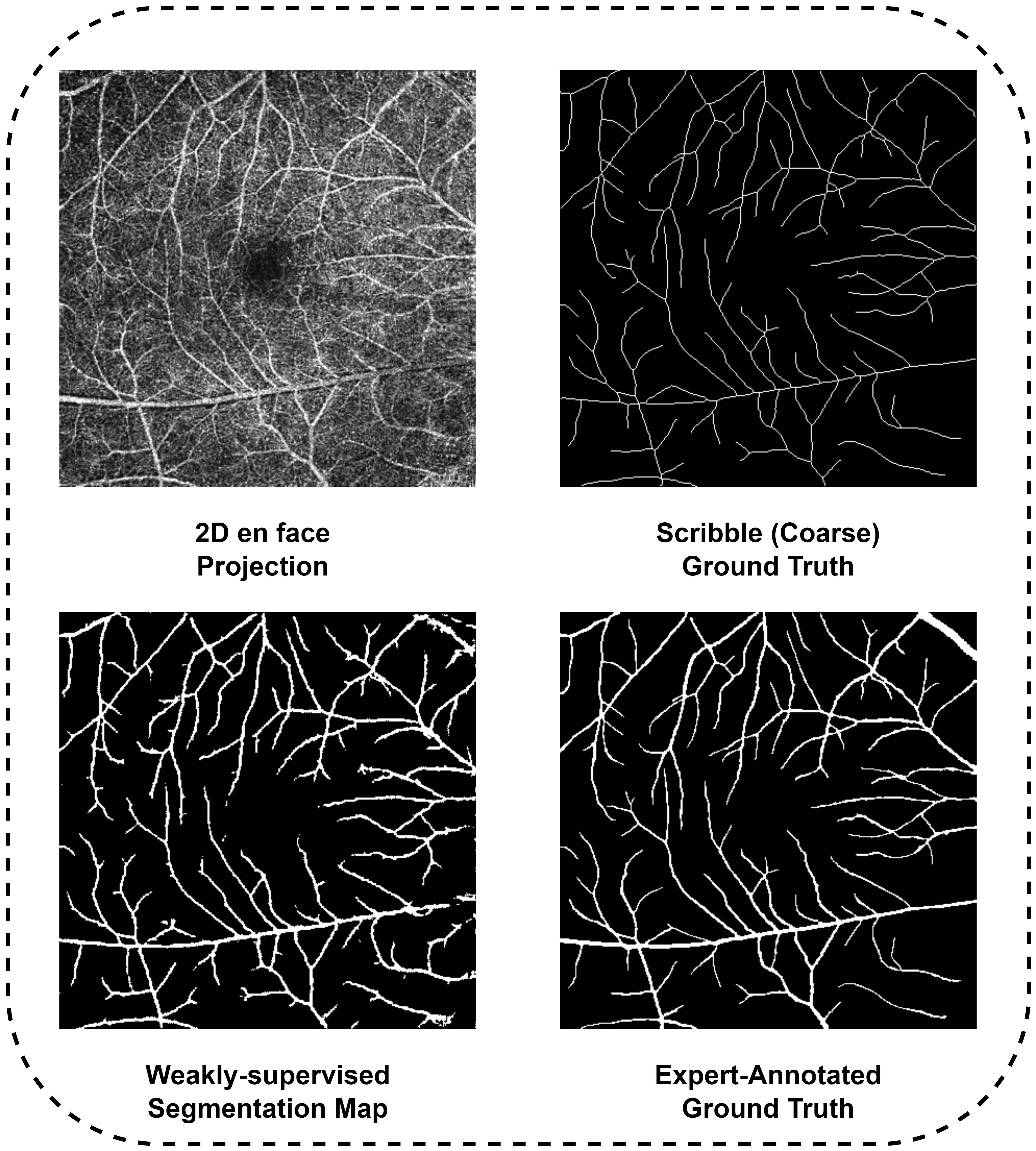}
    \caption{
    Alternative vessel segmentation output learned from the coarse annotation while achieving a limited loss in the segmentation performance.
    }
    \label{fig:intro_data}
\end{figure}

\IEEEPARstart{T}{he} visualization of retinal microvasculature has been an important ongoing research problem.
The effective visualization can help experts reveal   details that are useful for patient therapy. Many current researches focus on
 the techniques that can be used to improve the effectiveness of retinal vessel visualization.

 One of the techniques, called
  Optical Coherence Tomography Angiography (OCTA), a non-invasive imaging technique, has been shown to be helpful in the diagnosis of Diabetic Retinopathy (DR) and is a good alternative to the invasive Fluorescein Angiography \cite{Campbell2017_retinal_vascular_morphology, Tey2019_octa_dr_crreview, Sun2020_octa_dr_review, hormel_artificial_2021}. In the recent work, Ma \textit{et al}. discovered that the fractal dimension obtained from the 2D \textit{en face} projection of the microvascular structure within the parafovea area from the OCTA imaging technique has the potential to be used as an auxiliary biomarker for the Alzheimer's Disease (AD)\cite{ma_rose_2021}. However, such OCTA technique is prone to various noises and artifacts, such as shadow and projection artifacts. Thus, the vessel segmentation map obtained from the OCTA may result in the false detection of the vascular structure, which can lead to the  misdiagnosis.

Recent researches attempted to extract the vessel morphology  from either 2D or 3D OCTA images in order to provide an accurate representation of the retinal vasculature with both conventional \cite{Eladawi2017_DR_Handcraft_Automated, Zhao2017_Intensity_DR_Salient, Zhang2020_3dmodel_octa_dr} and deep learning methods \cite{gu_ce-net_2019, giarratano_automated_2020, li_ipn-v2_2020, mou_cs2-net_2021, ma_rose_2021}.  The conventional methods i.e., filters, and thresholding. These methods require the manual \textcolor{black}{calibration} to be perform \textcolor{black}{due to various capturing devices, and may also fail under certain conditions.} The processed outputs are also susceptible to noises and artifacts presented in the OCTA images.
For the deep learning methods, the manual calibration is not required and the resulting model can be generalized when  there are enough data for training. \textcolor{black}{While
not requiring the manual calibration, a large amount of data need to be annotated by experts in order to perform the fully-supervised training}.



Although, there exist many deep learning approaches on the task of vessel segmentation in OCTA images, to the best of our knowledge, only the work by Xu \textit{et al}. \cite{Xu2021_Patch_partial} attempted to mitigate the tenuous work in the labeling process of retinal vasculature in OCTA and Fundus images. The authors proposed the semi-supervised learning scheme that only requires some patches of label on the meaningful and informative area of the images. 
\textcolor{black}{In this approach, the need to fully-annotate the vessel for a patch and to determine which part of the image is actually contained the most useful information is still required.}

\textcolor{black}{In an attempt to rectify the expensive annotation process on the task} of the retinal vasculature for the 2D \textit{en face} OCTA images, we investigate the possibility of applying the scribble-like annotation \textcolor{black}{which can reduce the workload of experts dramatically since it does not require an expert-level precision in the annotation process. It is recently been used on the various other image segmentation tasks} on both medical and non-medical images \cite{lin_scribblesup_2016, wang_boundary_2019, Can2018_scribble_alone, ji_scribble-based_2019, valvano_learning_2021, Liu2022}. In particular, we propose a novel weakly-supervised learning framework for vessel segmentation tasks on the en face OCTA images, called \emph{OCTAve}. By enabling the usage of the scribble-like annotation as a supervised label to be used with the unpair, expert fully-annotated ground truth from another dataset can enhance the model learning by the adversarial game between the Segmentor and Discriminator network \cite{valvano_learning_2021}. The Segmentor training is constrained by our novel deep supervision mechanism called \textbf{Self-Supervised Deep Supervision (SSDS)}, which can significantly increase the model segmentation performance without requiring additional data (Fig. \ref{fig:intro_data}).
\textcolor{black}{Our contributions to the field of study can be listed as follows:}

\begin{itemize}
    \item We propose a method that can rectify the expensive labelling cost in 2D \textit{en face} OCTA images by incorporating the weakly-supervised learning method which reduces the need of fully-annotated labels.
    \item We propose a novel training mechanism called  \textit{Self-Supervised Deep Supervision}   used in the training of weakly-supervised model along with the adversarial deep supervision mechanism. \textcolor{black}{The results in Table \ref{table:1} and \ref{table:2} shows that our method significantly increased segmentation performance} in both the weakly-supervised and the fully-supervised settings.
    \item \textcolor{black}{To the best of our knowledge, our work is} the first in the field of retinal vessel analysis \textcolor{black}{which applies} the weakly-supervised learning together with the adversarial and self-supervised training on the deep neural nets.

\end{itemize}

\section{Related Works}
\textcolor{black}{In this section, we provide brief reviews of prior work on segmentation tasks using OCTA pictures and the prior work on the applications of several weakly supervised learning algorithms in medical imaging. In order to show the progress made in the tasks explored with OCTA image data, and the possible application of existing weakly-supervised learning technique to rectify our aforementioned problem introduced in the previous section.}

\subsection{Segmentation Tasks in OCTA}

Numerous works investigated the vessel segmentation tasks and the area segmentation tasks such as foveal avascular zone segmentation \cite{Mirshahi2021_faz_segmentation_dnn, Jabour2021_robust_faz_dnn, li_ipn-v2_2020}, by using deep neural networks on \emph{en face} OCTA images due to their performance being significantly higher and more robust than that of handcrafted filters \cite{hormel_artificial_2021}. L. Mou \emph{et al}. addressed the problem of curvilinear structure segmentation by proposing $\text{CS}^2$-Net \cite{mou_cs2-net_2021}, which incorporates a Channel and Spatial Attention Module into a U-Net-like autoencoder, resulted in the superior accuracy compared to the commonly used medical image segmentation algorithms such as U-net \cite{ronneberger_u-net_2015}, U-net++ \cite{zhou_unet_2020}, Attention U-net \cite{oktay_attention_2018}, R2U-net \cite{alom_recurrent_2018} and CE-net \cite{gu_ce-net_2019}. Li \emph{et al}. published the OCTA-500 dataset, which included pixel-level labeling for large retinal vessel and foveal avascular zone segmentation by proposed the novel 3D to 2D segmentation model IPN-V2, which takes 3D volumetric OCTA as an input and generated a 2D segmentation map \cite{li_ipn-v2_2020}. 

In the recent study, Ma \emph{et al}. published the ROSE dataset which contains \emph{en face} OCTA images with fully annotated vessel labels, as well as their proposed state-of-the-art architecture, OCTA-Net \cite{ma_rose_2021}, a novel coarse-to-fine network that employs detailed pixel-level annotation from a consensus of several experts and vessel centerline annotation to train a dual U-Net-like architecture with shared weights on the encoder stack in an attempt to aid the model learning the vascular structure in a joint-learning manner.

However, in these studies on the segmentation tasks, none had considered a way to mitigate the work that needs to be done by the experts in the labeling process despite the laborious work of the pixel-level annotation.

\subsection{Application of Weakly-supervise Learning in Medical Imaging}

Weakly-supervised learning has been in a focus of research in the deep learning field, with the goal of reducing the workload for the expert labeling while concurrently achieving an outstanding performance.

Jia \emph{et al}. proposed a weakly-supervised method for cancerous zone segmentation from histopathology images with the image-level annotation and constrained learning, which outperformed state-of-the-art architectures on a large-scale dataset illustrates the reducing amount of annotation work required \cite{jia_constrained_2017}. Fries \emph{et al}. tackled the task of classification of the aortic valve malformations from cardiac MRI images by using weakly supervision to generate noisy MRI labels and illustrated higher scores in all metrics compares to fully-supervised model on manual-labelled data \cite{fries_weakly_2019}. Xing \emph{et al}. approached the task of central serous chorioretinopathy (CSC) segmentation by two-stages learning architectures for weakly-supervision with image-level-only annotation, which greatly reduced the amount of labeling task \cite{xing_weakly_2021}.

Vepa \emph{et al.} proposed an automated cerebral vascular segmentation by using active contour as a weak annotation generator.  Their results illustrated slight lower scores but significantly reducing annotation time comparing to manual labeling for weak label \cite{Vepa2022}. Gondal \emph{et al}. achieved high accuracy, low false positives with high sensitivity in detecting lesion region in retinal images by using lesion-level and image-level annotation for weakly-supervised boundary localization, achieving commensurate or even better performance than fully-supervise method \cite{Gondal2017}. Liu \emph{et al}. used scribble annotation enhanced with uncertainty-aware self-ensembling and transformation-consistent approaches for weakly-supervised COVID19 infectious area segmentation task from CT images; the result on several data sets presents higher efficiency than other weakly-supervised method while obtaining similar performance to fully-supervised method \cite{Liu2022}.

The above weakly-supervised medical imaging approaches all show the performance improvement and manual labelling workload decrement. Thus, it is worthwhile to investigate the use of weakly supervised learning for medical image-related tasks, primarily to reduce the laborious work required of the human experts and to increase data availability. In this paper, we found that the  scribble-like ground truth can be used to mitigate the previously described issues with the weakly-supervised technique for 2D \textit{en face} OCTA vessel segmentation.

\section{Methods}

\begin{figure*}[t]
    \centering
    \includegraphics[width=\textwidth]{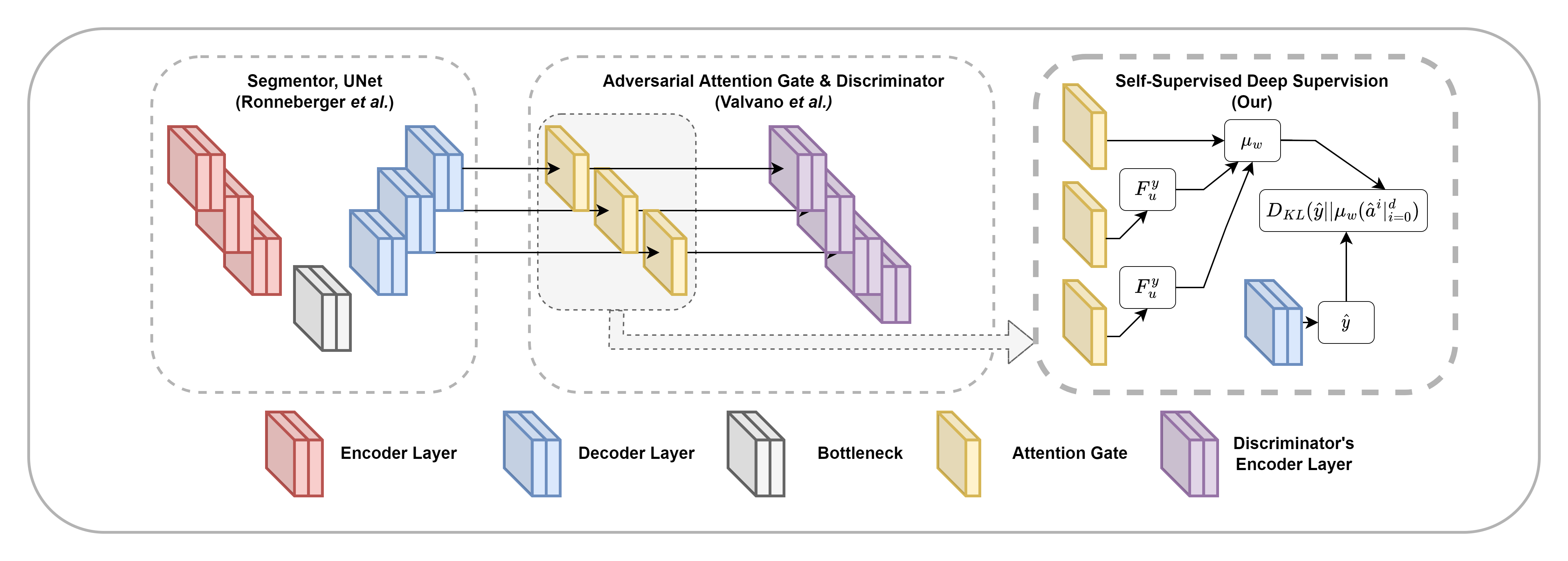}
    \caption{The proposed architecture overview of OCTAve.  It consists  of 3 parts, the UNet-like Segmentor (Left), Adversarial Attention Gates and the Discriminator (Center), and the Self-Supervised Deep Supervision Mechanism (Right).  \textcolor{black}{Each part has different learning objectives but has a common goal of vessel extraction.}}
    \label{fig:arch1}
\end{figure*}

Our proposed method is illustrated in Fig. \ref{fig:arch1}. \textcolor{black}{The network architecture} adopts the work of Valvano \textit{et al} \cite{valvano_learning_2021} and significantly augmented by our novel self-supervised deep supervision mechanism as presented in the Section \ref{ssds}.
 
\begin{figure}[]
    \centering
    \includegraphics[width=\columnwidth]{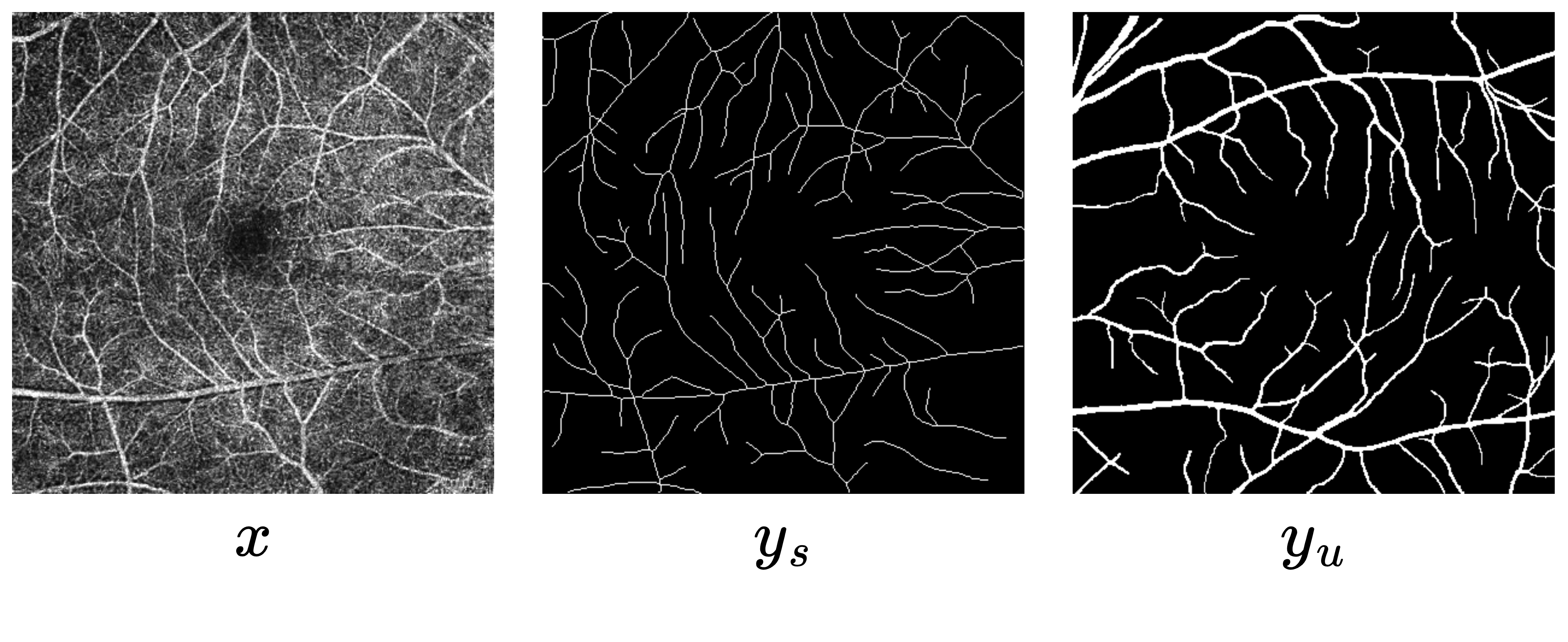}
    \caption{Example of the training data used to train the Segmentor ($x, y_s$) and the Discriminator ($y_u$) from OCTA-500 6M.}
    \label{fig:training_sample_ex}
\end{figure}

The approach assumes the existence of a freely accessible, expert-labeled dataset.
The dataset may be indirectly-related to our dataset. For example, images may be retrieved from the different devices, but the underlying ground truth is similar to our target dataset; we called this dataset an unpaired dataset. \textcolor{black}{Examples of the data are shown in} Fig.\ref{fig:training_sample_ex}.
To utilize the available expert-made annotations to help in the training of weakly labeled datasets,   the concept of domain transfers using a generative adversarial network (GAN) \cite{goodfellow2014generative} is deployed. The Segmentor network  attempts to fool the Discriminator network which attempts to judge the difference between the expert-made segmentation map of the unpaired dataset and the model-predicted segmentation map of the targeted dataset. \textcolor{black}{The goal is to make  the Segmentor learn about the shape prior from the unpair expert-annotated ground truth}. Thus, the adversarial game played between the Segmentor and the Discriminator enabling the Segmentor to produce a segmentation map resembling to the expert-annotated ground truth.

Let us denote $\Sigma$ and $\Delta$ as the Segmentor and the Discriminator network respectively. We assign $x$, $y_{s}$ and $y_{u}$  as an input, a scribble-like ground truth, and a unpair fully-annotated ground truth respectively, \textcolor{black}{where $\hat{y}, \hat{a}^{i}|^{d}_{i=0}$ and $c$ are a predicted segmentation map, a set of attention maps and the discriminator output. The functional of each part can be described into equations as follow}:
\begin{equation}
    \begin{split}
    \Sigma(x) = \hat{y}, \hat{a}^{i}|^{d}_{i=0}
    \end{split}
\end{equation}
\begin{equation}
    \begin{split}
    \Delta(\hat{a}^{i}|^{d}_{i=0}) = c; c \in \mathbb{R}
    \end{split}
\end{equation}

\begin{figure}[]
    \centering
    \includegraphics[width=\columnwidth]{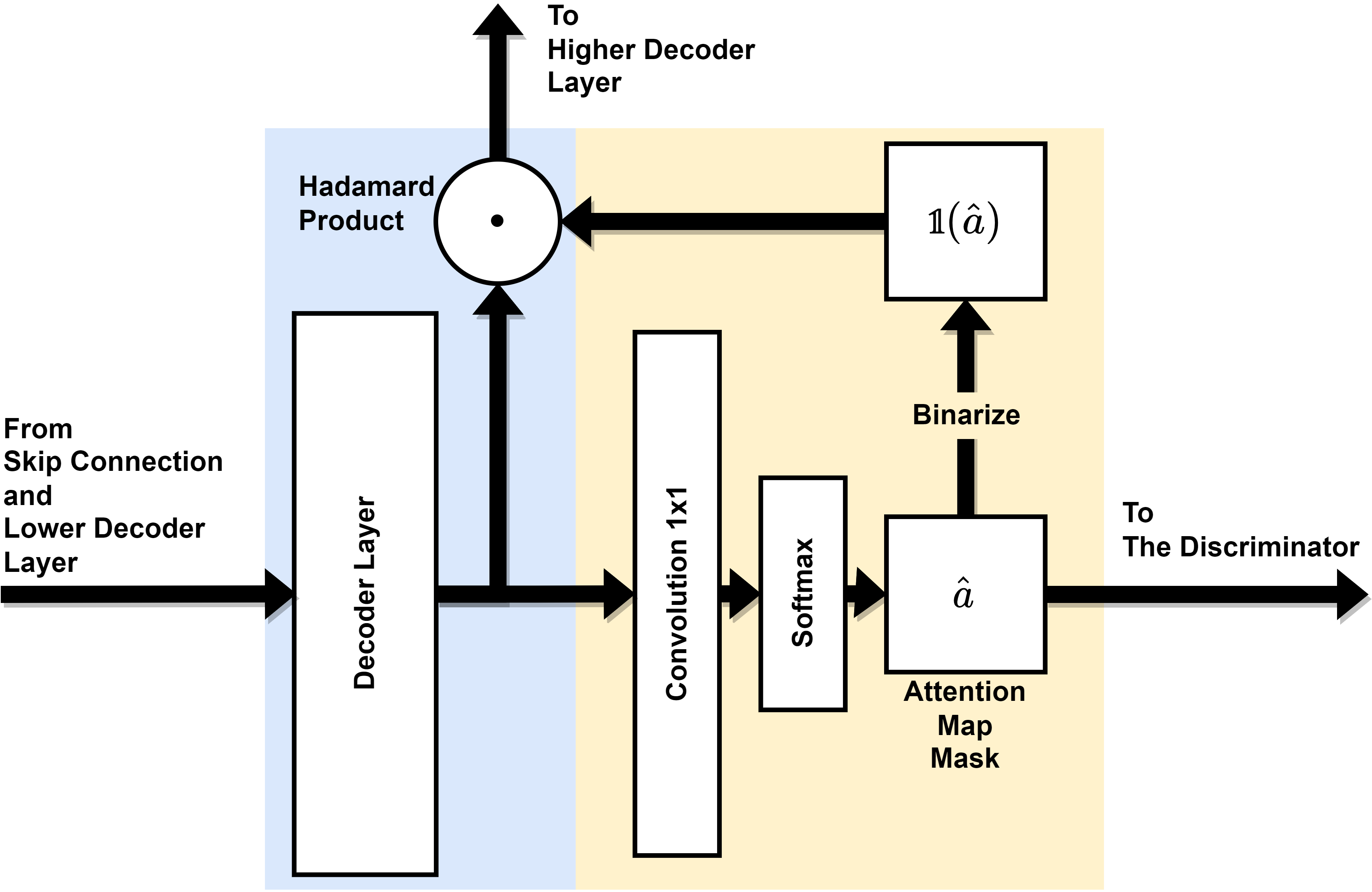}
    \caption{Adversarial Attention Gate architecture, a part of adversarial deep supervision mechanism \cite{valvano_learning_2021}.}
    \label{fig:aag}
\end{figure}

The network is trained to   fool the Discriminator at multiple resolutions, by exploiting the stack of decoder layers. The adversarial game is played at the discriminative outputs from each decoder layer. Then, the outputs are fed to the adversarial attention gates as 
in Fig. \ref{fig:aag} before passing through the discriminators.
The adversarial game optimization at  multi-scale outputs of the UNet are named as \emph{adversarial deep supervision}.

In the above equations, $\hat{a}^{i} |^{d}_{i} = \{\hat{a}^0, \hat{a}^1, ..., \hat{a}^d\}$ is a set of attention map outputs from the adversarial attention gates, where $d$ is the depth of the segmentor's attention gates.

\begin{equation}\label{mwpce}
    \begin{split}
        & W_{pce} = \mathds{1}(y_s)[-\sum^{c}_{i=1}w_iy_{s_i}\log({\hat{y}_i})]
    \end{split}
\end{equation}
\begin{equation}\label{lsganloss}
    \begin{split}
        \mathcal{V}_{\Sigma} = & \frac{1}{2}E_{x\sim X}[(\Delta(\Sigma(x)) - 1)^2] \\
        \mathcal{V}_{\Delta} = & \frac{1}{2}E_{x\sim X}[(\Delta(\Sigma(x)) + 1)^2] \\
        & + \frac{1}{2}E_{y\sim Y}[(\Delta(y_{u} - 1)^2]
    \end{split}
\end{equation}

For the supervised loss, masked version of weighted partial cross entropy introduced in \cite{valvano_learning_2021} denoted in  Equation \ref{mwpce} are used for weakly-supervised learning.  The weighted term $w_i$ is the ratio between the number of annotated pixels of class $i$ over the number of annotated pixels of every classes, including the background. The least-square GAN losses for both the Segmentor and the Discriminator are denoted in Equation \ref{lsganloss}. LS-GAN loss is used for an adversarial game optimization due to its effectiveness in the style adaptation tasks, which is suitable for our  training purpose; ie., the Segmentor  adapts to the expert annotation style in its prediction. On the other hand, the Segmentor loss $\mathcal{L}_{\Sigma}$ and the Discriminator loss $\mathcal{L}_{\Delta}$ can be denoted as follows:
\begin{equation}
    \begin{split}\label{model_loss}
    \mathcal{L}_{\Sigma} = &\ \alpha_0 W_{pce}(\hat{y}, y_{s}) + \alpha_1 \mathcal{V}_{\Sigma}(\hat{y}, \hat{a}_i|^{d}_{i=0}) \\
    \mathcal{L}_{\Delta} = & \ \alpha_2\mathcal{V}_{\Delta}(\hat{a}_i, F^{\hat{a}_i}_{d}(y_{u}))    
    \end{split}
\end{equation}

In Equation \ref{model_loss},
$\alpha_0$ is a dynamic weight used to balance the optimization between weakly-supervised learning and adversarial game optimization. Note that, in our approach,
We will modify this term  to achieve  
the faster training and more stability in the large parameter model. This   will be discussed later in the Section \ref{sectionIIIsubsectionA}.
Lastly, $\alpha_1$ and $\alpha_2$ are the fixed weights that are regulated the adversarial loss optimization of both the Segmentor and the Discriminator, which have been empirically set to 0.1 according to the setting as in the original work.

\subsection{Balancing Mechanism of Multi-objective Learning}
\label{sectionIIIsubsectionA}

\textcolor{black}{It is well-known that maintaining the training stability of the adversarial neural network is a difficult task, and it is even more difficult for multi-objective learning like our architecture. 
To achieve the stability, by avoiding the over-optimization on the only one of the objectives, an adaptive strategy is a must.} Herein, in this section, we will discuss about the original dynamic weight term and our proposed modification.

From $\mathcal{L}_{\Sigma}$ in the Equation \ref{model_loss}, $\alpha_0 = \frac{||\mathcal{V}_{\Sigma}||}{||W_{pce}||}$ is a dynamic weight term, defined by the ratio between adversarial loss and supervised loss with an intention to keep the optimization balance between weakly-supervised objective and adversarial game. However, while this method allows stable training, the model training  process takes a long time to reach the optimal point.

In this work, \textcolor{black}{the preliminary study on} the trade-off of the training stability for the faster descent \textcolor{black}{were conducted and observed}. We introduce an alternative reciprocal version of the dynamic weight term as in Equation \ref{alternateAlpha0}.
\begin{equation}
\label{alternateAlpha0}
    \alpha_0 = 
    \text{Clamp}(\frac{||W_{pce}||}{||\mathcal{V}_{\Sigma}||}), \text{max}=\mathcal{C})
\end{equation}

In Equation \ref{alternateAlpha0}, $\mathcal{C}$ is a constant that prevents the model from overly attending to weakly-supervised learning optimization during the early training stage. We have empirically set the value to 0.1 based on the observation from the behavior during the model training.  To show the effectiveness of our proposed modification,  Section \ref{effect of alpha0} shows the experimental results.

\subsection{Self-supervised Deep Supervision}
\label{ssds}

\begin{figure}
    \centering
    \includegraphics[width=\columnwidth]{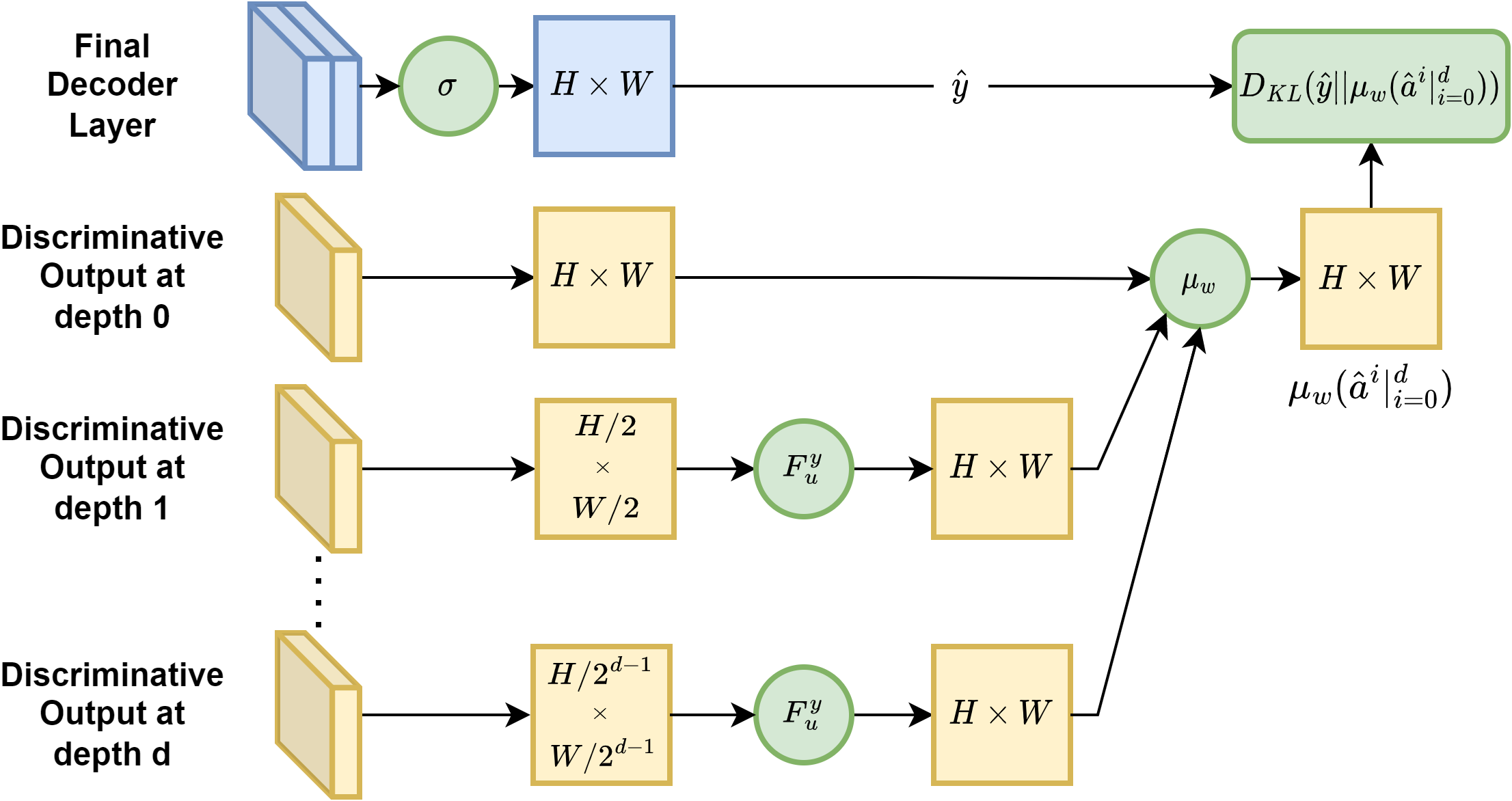}
    \caption{Our proposed self-supervised deep supervision working mechanism for any N-depth UNet-like architecture with discriminative output on the decoder layer. $\sigma, F^y_u, \mu_w$ denote softmax function, upscaling function and weighted average function respectively.}
    \label{fig:ssds}
\end{figure}

As one of the main contribution of our study, we propose a novel mechanism for deeply supervise training on a UNet\cite{ronneberger_u-net_2015} like architecture called \textbf{Self-Supervised Deep Supervision} (Fig. \ref{fig:ssds}). This mechanism deeply supervises the model during the training based on the pixel-wise confidence of the segmentation map. Based on the observation   by   Zhao \textit{et al}. \cite{zhao_dsal_2021}, a well-performed UNet model with discriminative output at each of the decoder layers tended to have high consistency of Dice's coefficient between the discriminative output from each decoder layer. In our case, the attention maps from the adversarial attention gates are also discriminative output from the decoders. This led us to the idea of increasing the consistency between decoder layers' discriminative outputs as one of the optimization objectives.

In other words, the decoder layer can be thought of as a probability density estimator function that is parameterized by its own weight $\theta$.
\begin{equation}
 Decoder_i(x;\theta_i) \equiv PDE(x;\theta_i)
\end{equation}
From the previous statement, we can argue that each decoder layer should be mapped to the approximately same distribution in order to produce highly consistent and higher precision segmentation maps.
\begin{equation}
    PDE(x; \theta_d) \sim PDE(x; \theta_i);\text{ } 0 \leq i < d
\end{equation}
In our case, using Dice's coefficient as a measurement metric is not an option due to the usage of weak annotation, and in some cases, we may not have labels for some of the images. Instead, we incorporate the Kullback-Leibler Divergence \cite{Kullback1951_KL} to improve the consistency between attention maps from each layer. We can say that our objective is to minimize the divergence of confidence between attention maps of each layer. The objective function can be denoted as follows:
\begin{equation}
    \text{minimize} \sum_{i=0}^{d} D_{KL}(\hat{y} || Decoder_i)
\end{equation}
Thus, we call the application of the deep supervision based on \emph{KL} divergence loss function as Inter-layer Divergence Loss (ILD) as follows:
\begin{equation}
    \begin{split}
    \mathcal{L}_{ILD}(\hat{y}, \hat{a}_{i}|^{d}_{i=0})
    & = \hat{y} [\log(\hat{y}) - \log(\mu_w(\hat{a}_{i}|^{d}_{i=0})) ]
    \end{split}
\end{equation}
 In the above equation, $\hat{y}$ is the predicate output from the Segmentor which we use as a posterior, and $\hat{a}_{i}|^{d}_{i=0}$ is a set of the attention map outputs from the adversarial attention gates.
\begin{equation}
\label{mu_w}
\mu_w(\hat{a}_{i})|^{d}_{i=0} = \frac{1}{d}\sum^{d}_{i=0} w_i F^{y}_{u}(\hat{a}_i)
\end{equation}
Instead of pairwise calculation between posterior and each of the attention maps, weighted average confidence from the set of attention maps specified in Equation \ref{mu_w} are used for the implementation of this work,  to reduce the computational cost from the pairwise calculation of the divergence between each layer. $F^{y}_{u}$ is the differentiable upscaling function used to upscale the lower resolution attention maps to match the posterior resolution.

Thus, we can rewrite the segmentor loss function with $\mathcal{L}_{ILD}$ as an additional optimization objective as follows:
\begin{equation}
    \mathcal{L}_{\Sigma} = \alpha_0 W_{pce}(\hat{y}, y_{s}) + \alpha_1 \mathcal{V}_{\Sigma}(\hat{y}) + \kappa \mathcal{L}_{ILD}(\hat{y}, \hat{a}_{i}|^{d}_{i=0})
\end{equation}

\begin{equation}\label{kappa_term}
    \kappa = \frac{||\mathcal{L}_{ILD}||}{||W_{pce}||+||\mathcal{V}_{\Sigma}||}
\end{equation}

Equation \ref{kappa_term} is an additional dynamic weight  introduced to keep the balance of optimization between self-supervised learning and others.

\section{Experiments \& Results}

\begin{figure*}[]
    \centering
    \includegraphics[width=\textwidth]{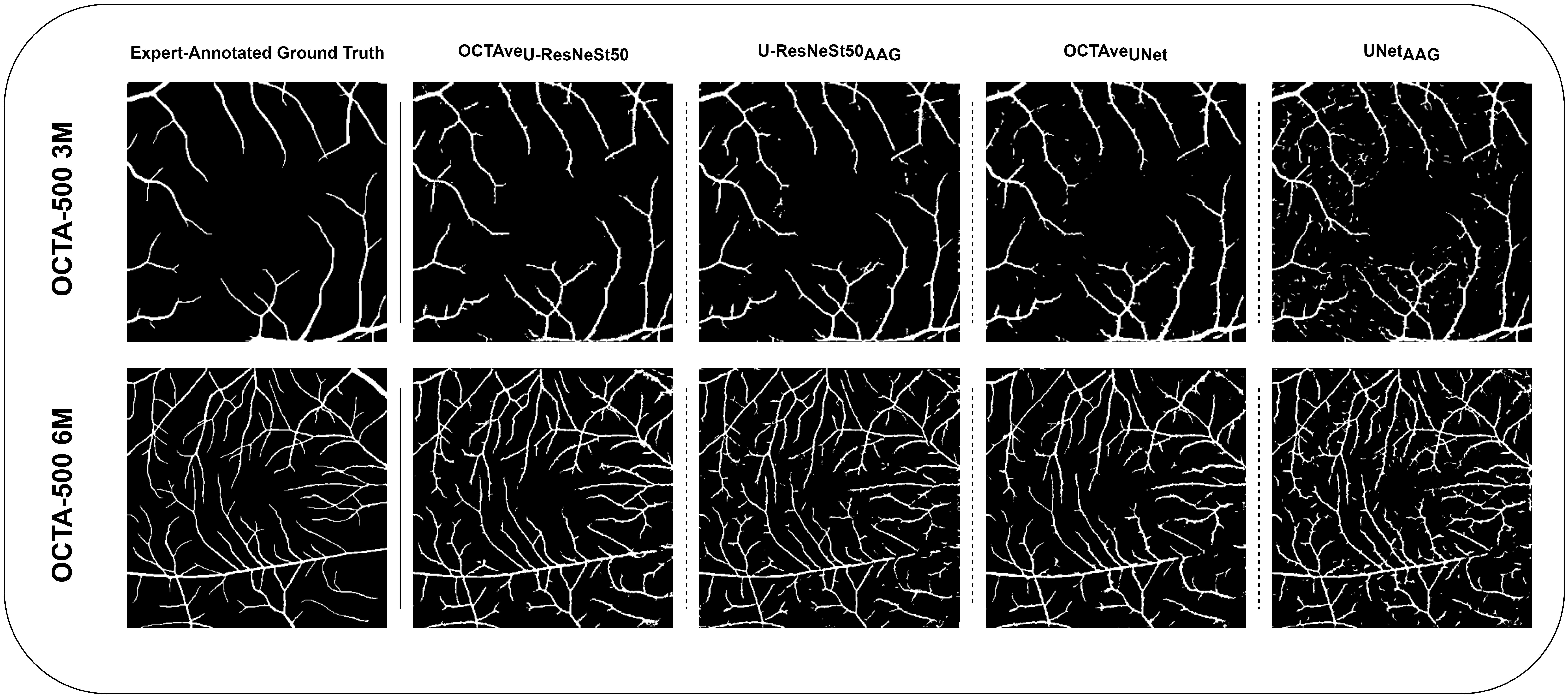}
    \caption{Segmentation map of the sample from test set of the OCTA-500 3M and 6M on each methods in  Section \ref{ExIBenchmark}. The segmentation map were ordered accordingly, from left to right: expert-annotated ground truth, segmentation map from the inference of the 2D en face angiogram through OCTAve\textsubscript{U-ResNeSt50}, U-ResNeSt50, OCTAve\textsubscript{UNet}, UNet.}
    \label{fig:result_viz}
\end{figure*}

Our study consists of two main experiments to show the effectiveness of the proposed method on the task of vessel extraction of 2D \textit{en face} OCTA images. The weakly-supervised training experiment in the Section \ref{ExperimentI} were conducted as our main contribution of this study. Then, to show the useful application on other than weakly-supervised learning of our proposed method, the fully-supervised training experiment were conducted as presented in the Section \ref{ExperimentII}.

\subsection{Experiment I: Weakly-supervised Segmentation}
\label{ExperimentI}

\begin{table*}[h]
\resizebox{\textwidth}{!}{
\ra{1.3}\begin{tabular}{@{}ccccccccc@{}}
\toprule[1.2pt]
 \multirow{2}{*}{\textbf{Modality}} & \multirow{2}{*}{\textbf{Projection}} & \multirow{2}{*}{\textbf{Scribble \%}} & \multicolumn{6}{c}{\textbf{Method}}                                                          \\ & & & \textbf{OCTA-Net} & \multicolumn{2}{c}{\textbf{U-ResNeSt50}} & \multicolumn{2}{c}{\textbf{UNet}} & \textbf{OCTA-Net}
                         \\
                         \cmidrule(l){4-4}
                         \cmidrule(l){5-6}
                         \cmidrule(l){7-8}
                         \cmidrule(l){9-9}
                          & & & \textbf{OCTA-Net\textsubscript{UB}} & \textbf{U-ResNeSt50\textsubscript{AAG}} & \textbf{OCTAve\textsubscript{U-ResNeSt50}} & \textbf{UNet\textsubscript{AAG}} & \textbf{OCTAve\textsubscript{UNet}} & \textbf{OCTA-Net\textsubscript{LB}} \\ 
                          \cmidrule{1-9}
\multirow{9}{*}{3M}       & \multirow{3}{*}{FULL}  & 100\% & 87.83      & 76.94         & \textbf{78.48*}                             & 72.99  & \textbf{76.94*}  & 42.06
                         \\ &&75\%&&75.72&\textbf{76.07*}&73.26&\textbf{74.66*}&
                         \\ &&50\%&&74.84&\textbf{75.44}&71.23&\textbf{72.43*}&
                         \\ &&10\%&&\textbf{70.66*}&67.16&\textbf{69.16*}&66.45&
                        \\ \cmidrule{2-9}
                           & \multirow{3}{*}{ILM-OPL} & 100\%                     & 91.32      & 81.42         & \textbf{82.35*}              & 77.20  & \textbf{82.34*}       & 53.13
                        \\ &&75\%&&\textbf{81.38}&80.89&80.97&\textbf{81.08}&
                        \\ &&50\%&&80.31&\textbf{80.67}&80.18&\textbf{80.52}&
                        \\ &&10\%&&\textbf{77.37*}&75.41&\textbf{78.94*}&77.87&
                           \\ \cmidrule{2-9} & \multirow{3}{*}{OPL-BM} & 100\%                     & 82.62      & 71.07         & \textbf{73.86*}              & 68.07  & \textbf{72.36*}       & 26.05
                          \\ &&75\%&&\textbf{72.88*}&{71.95}&68.86&\textbf{70.52*}&
                          \\ &&50\%&&70.20&\textbf{70.52}&67.14&\textbf{67.68}&
                          \\ &&10\%&&\textbf{67.58*}&65.26&64.71&\textbf{64.97}&
                         \\ \cmidrule{1-9}
\multirow{9}{*}{6M} & \multirow{3}{*}{FULL} & 100\% & 84.62      & 74.33         & \textbf{77.17*}              & 72.71  & \textbf{75.26*}       & 38.33
                        \\ && 75\%&&73.82&\textbf{76.45*}&69.49&\textbf{74.72*}&
                        \\ && 50\%&&73.24&\textbf{73.93*}&69.29&\textbf{70.68*}&
                        \\ &&10 \%&&\textbf{71.03*}&70.48&\textbf{70.02*}&68.79
                         \\ \cmidrule{2-9}
                          & \multirow{3}{*}{ILM-OPL} & 100\%                     & 88.87      & 79.48         & \textbf{81.43*}              & 78.76  & \textbf{81.06*}       & 46.23
                        \\ && 75\%&&\textbf{79.19}&{78.23}&78.61&\textbf{80.96*}&
                        \\ &&50\%&&78.83&\textbf{79.15}&78.61&\textbf{80.57*}&
                        \\ &&10\%&&\textbf{78.63*}&77.62&77.97&\textbf{78.46*}&
                        \\ \cmidrule{2-9}
                          & \multirow{3}{*}{OPL-BM} & 100\%                     & 78.84      & 68.56         & \textbf{72.16*}              & 65.06  & \textbf{69.39*}       & 35.59
                          \\ && 75\%&&70.15&\textbf{71.85*}&66.63&\textbf{69.48*}&
                          \\ &&50\% &&67.35&\textbf{67.55}&64.31&\textbf{66.46*}&
                          \\ &&10\% &&\textbf{64.68*}&63.34&\textbf{63.77*}&62.98&
                        \\ \bottomrule[1.2pt]
\end{tabular}
}
\captionsetup{width=\textwidth}
\caption{Vessel segmentation performance (Dice's Coefficient) of the 100\%, 75\%, 50\% and 10\% scribble availability variation compared across the upper bound, lower bound and methods. Bold text denotes the method with the best numerical value within the same architecture, and * denotes if the method is statistically significant ($p < 0.001$).}
\label{table:1}
\end{table*}

\subsubsection{Datasets and Data Preparation}

\begin{figure}
    \centering
    \includegraphics[width=\columnwidth]{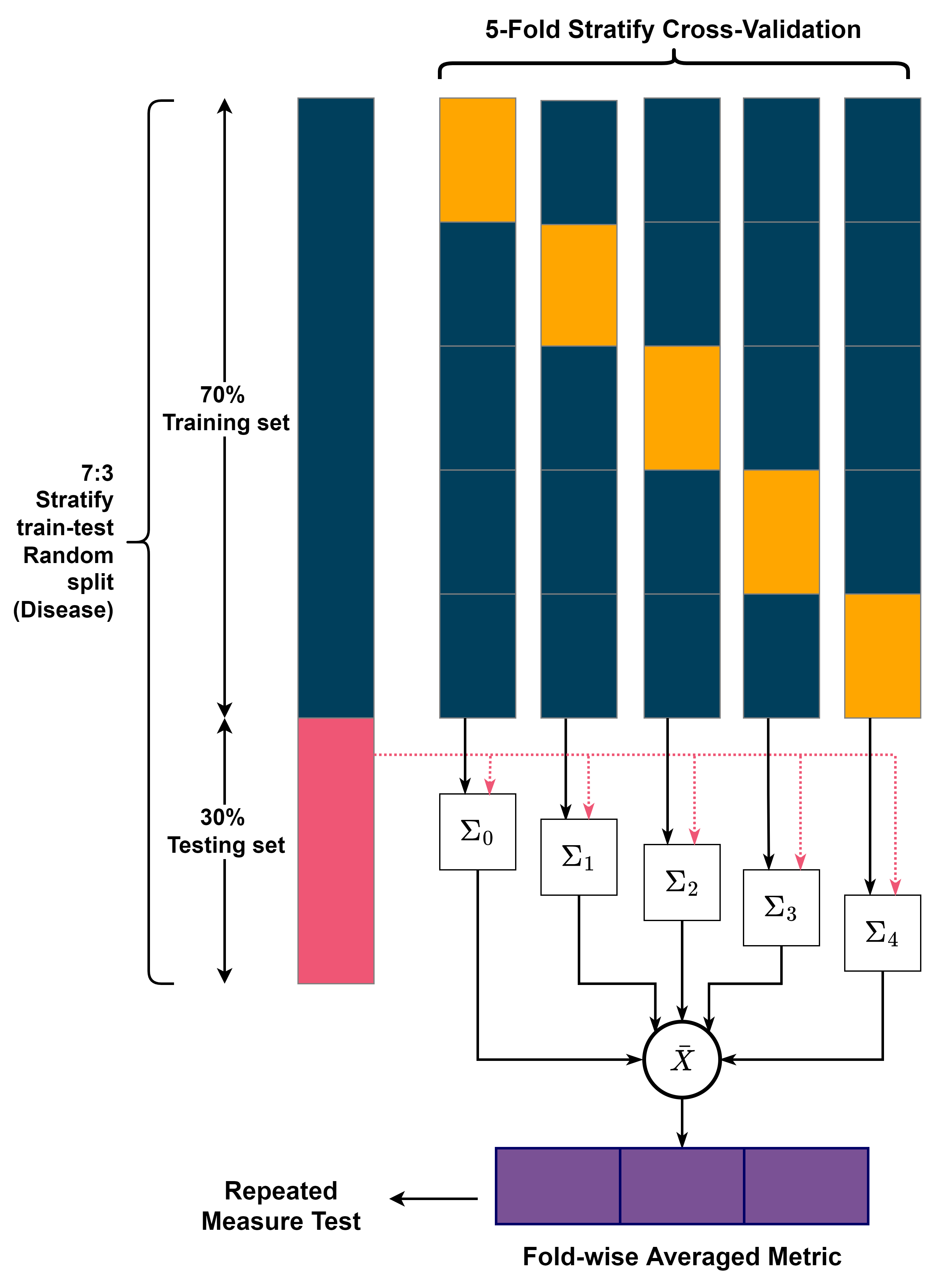}
    \caption{Experiment I data preparation and evaluation. $\Sigma_i$ denote the best validation score Segmentor model for fold $i$.}
    \label{fig:experimentIdata}
\end{figure}

In this experiment, we used the OCTA-500 dataset \cite{Li2019_octa500_dataset}, a public multi-modality dataset with total of 500 subjects. The dataset provides two types of field of view (FOV), a $3 \times 3\ cm^2$ FOV dubbed as OCTA-3M containing 200 subjects and a $6x6\ cm^2$ FOV dubbed as OCTA-6M containing 300 subjects. Each type of FOV contains three types of 2D \textit{en face} OCTA projection with different depths as follows: FULL, ILM-OPL, and OPL-BM projection. Text labels for diseases label are also provided for non-normal sample.

For this experiment, the training and the evaluation of each projection and FOV were performed separately. In the preparation of the dataset, a 70:30 stratify random train-test split based on the diseases label with random state seed of 50, followed by the 5-fold stratify cross validation on the train dataset were performed on each group. Additionally, an augmentation of the training set were done by randomly rotating the input image and its corresponding label within the range of -10 to 10 degree. In order to simulate the unpair dataset, we randomly selected half of the training set to be used as an unpair dataset.

Since the original OCTA-500 does not provide an expert-made scribble annotation, we synthetically created the scribble annotation to be used as a weakly supervised ground truth by performing the skeletonization of a pixel-level annotation provided by the dataset obtained by Zhang's algorithm \cite{Zhang1984AFP}.

Additionally, to explore our robustness and possible drawbacks of our method where the model relied on the adversarial knowledge for unsupervised sample and the enforcement of self consistency correction, which may cause the error to propagated during the training, the experimentation was variated by the scribble label available for the use in the training set as in \cite{valvano_learning_2021} as follows: 100\%, 75\%, 50\% and 10\% scribble availability.

\subsubsection{Benchmarking Methods}
\label{ExIBenchmark}
The validation of efficacy of automated segmentation methods was performed by comparing our proposed method against the current fully-supervised state-of-the-art method, OCTA-Net \cite{ma_rose_2021} in both weakly-supervised version as a lower bound baseline and fully-supervised version as a soft upper bound, and the current state-of-the-art scribble-base weakly supervise method from \cite{valvano_learning_2021}. We considered the following:

\begin{itemize}
    \item \textbf{OCTA-Net\textsubscript{UB}} \cite{ma_rose_2021}: OCTA-Net, the current state-of-the-art fully-supervised automated vessel segmentation method on 2D \textit{en face} OCTA images. Their method is a two-stage coarse-to-fine network. The coarse stage network is essentially a UNet \cite{ronneberger_u-net_2015} with the imagenet-pretrain ResNeSt50\cite{zhang_resnest_2020} as a backbone for the encoder layers and two decoder branch.
    \item  \textbf{OCTA-Net\textsubscript{LB}}: We considered the scenario where one tried to apply the scribble weakly-supervised learning objective to OCTA-Net. In this case, $W_{pce}$ (Equation \ref{mwpce}) loss was used for the fair comparison with our methods.
    \item \textbf{UNet\textsubscript{AAG}}: UNet with the adversarial attention gate and the Discriminator network from \cite{valvano_learning_2021}. The configuration of the architecture and the hyperparameters are the same as the original work, except for the dynamic weight $\alpha_0$, which was changed into an alternate version as proposed in the Equation \ref{alternateAlpha0}.
    \item \textbf{OCTAve\textsubscript{UNet}}: Based on \textbf{UNet\textsubscript{AAG}}, self-supervised deep supervision mechanism was incorporated into the training of the Segmentor network to enhance the segmentation performance and improves the consistency of attention maps between decoder layers.
    \item \textbf{U-ResNeSt50\textsubscript{AAG}} and \textbf{\textbf{OCTAve\textsubscript{U-ResNeSt50}}}: To show the effectiveness of our proposed method on the complex architecture with high number of parameters, the vanila UNet encoder that was used in both the \textbf{UNet\textsubscript{AAG}} and the \textbf{UNet\textsubscript{AAG-SSDS}}, which had far fewer number of weight parameters, was replaced with ResNest50, a large, complex architecture used in the coarse stage network of the OCTA-Net.
\end{itemize}

\subsubsection{Model Training and Evaluation Methods}

The model and the experimentation were implemented using PyTorch Lightning framework \cite{Falcon_PyTorch_Lightning_2019}. The global random state seed was set as 50, and the deterministic training option was enabled to maximize the reproducibility. The training epochs were set to 1,000 epochs along with the checkpointing with the best Dice's coefficient score on a validation dataset to be used a criterion for every benchmarking method. As for the optimizer, we use Adam optimizer \cite{Kingma_adam_2014} with the learning rate and weight decay parameters were set to 0.001 and 0.0001 respectively, and the cyclic learning rate scheduler were used with the oscillation between 0.0001 and 0.00001. Except for  \textbf{OCTA-net\textsubscript{UB}}, we used the same configuration as specified in \cite{ma_rose_2021}, e.g. polynomial learning rate scheduler with a power factor of 0.9.

The segmentation performance was reported using Dice's coefficient score, the \textcolor{black}{regularly} used evaluation metric in the image segmentation tasks. The statistical significance of the reported results was tested using non-parametric Friedman repeated measure, followed by the pairwise t-test to calculate statistical significant value with Bonferroni p-value correction. $p < 0.001$ is considered to be statistical significant.

\subsubsection{Performance Analysis}

\begin{figure}
    \centering
    \includegraphics[width=\columnwidth]{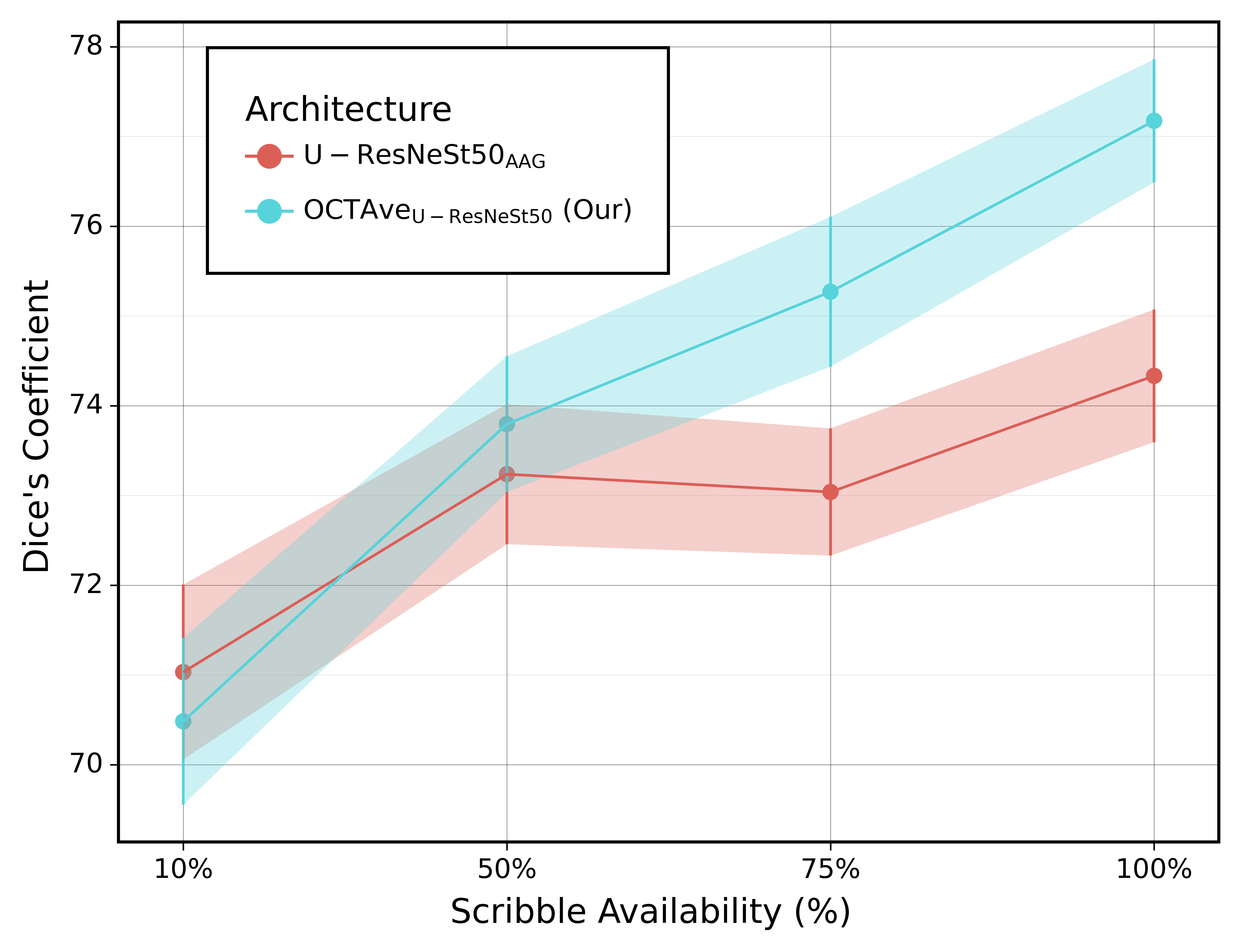}
    \caption{Segmentation performance comparison between \textbf{OCTAve\textsubscript{U-ResNeSt50}} and baseline \textbf{U-ResNeSt50} on OCTA-500 6M-FULL test set.}
    \label{fig:octanet6mcomp}
\end{figure}

The superiority of our proposed methods was verified by comparing the segmentation performance on the deep neural network architecture incorporated with our proposed self-supervised deep supervision mechanism against the one without reported in Table \ref{table:1}.

From the experiment, the major difference in the segmentation performance occurs at the application of our method and the variation of the scribble availability.  We summarize the discussions as follow:
\begin{itemize}
\item \textbf{Impact of the self-supervise deep supervision mechanism.} As the mechanism is the main contribution of this study, the comparison of our proposed method was conducted on both the lightweight vanilla UNet and the heavyweight ResNeSt50-backboned UNet to show the effectiveness of the application. In this part, we only consider the 100\% scribble availability variation to exclude any exposure to the unsupervised learning from the analysis.
 
Based on the results shown in Table \ref{table:1}, both \textbf{UNet} and \textbf{U-ResNeSt50} architectures, the segmentation performance between the one with our proposed method and the one without, was drastically different as the statistical analysis performed with the significant value of 0.001. With the improvement seen on the architectures incorporated with our proposed method on every group of projection level and FOV, we can argue that our method has a highly beneficial impact on the segmentation performance of the weakly-supervised learning setting of 2D \textit{en face} OCTA vessel segmentation.
  
Consequently, the experiments show the possible application of the weakly-supervised learning method using the scribble-like label annotation proposed in \cite{valvano_learning_2021} to achieve a limited reduction in the segmentation performance of curvilinear structure despite the usage of coarse labels, and significantly augmented by our proposed deep supervision method. Additionally, our method has potential to be applied for other segmentation tasks such as foveal avascular zone or other different organs, with improved segmentation quality.
 
\item \textbf{Robustness of the mechanism over the exposure to unsupervised learning.} To investigate the possible drawback of our proposed method in the exposure to the unsupervised learning when the label is missing, experiments on the 75\%, 50\% and the 10\% variation of scribble availability were considered. Experimentation results on the 50\% variation have shown that our method's difference in the segmentation performance were mostly statistically insignificant. While the results on the 10\% variation shown our method suffered from the error propagation and achieved lower segmentation performance. As a result, the experiments revealed that our method's robustness can withstood as high as 50\% of the label being missing before gradually outperformed by the unaugmented method.
\end{itemize}

\begin{figure}
    \centering
    \includegraphics[width=\columnwidth]{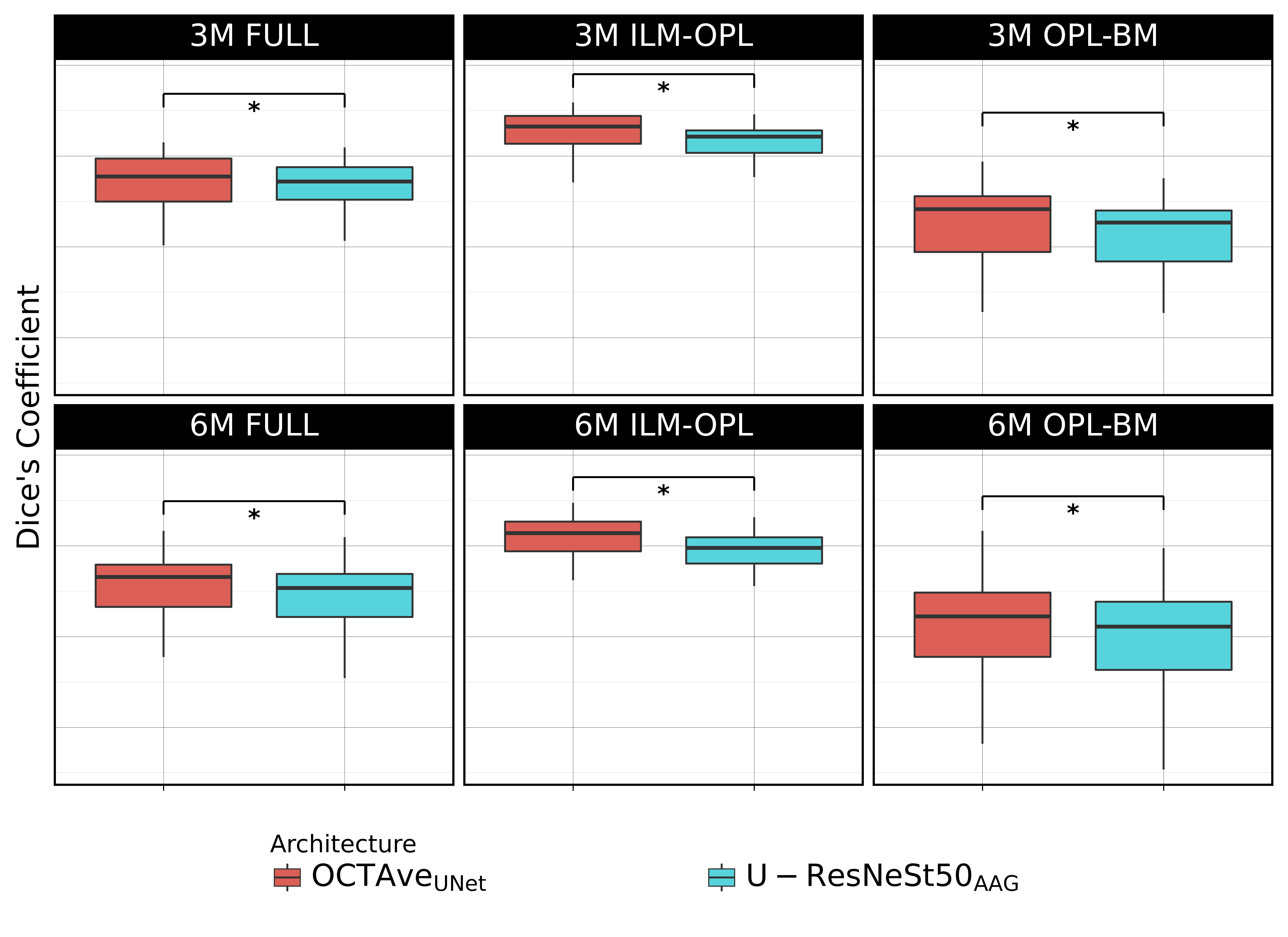}
    \caption{Performance comparison between the lightweight and the heavyweight \textbf{OCTAve\textsubscript{UNet}} against \textbf{U-ResNeSt50}. Demonstrating the significant competitive advantage of our proposed method in the enhancement of the low parameter network to outperform the large parameter network. * denote the statistically significance of $p<0.001$.}
    \label{fig:advcomp}
\end{figure}

\subsection{Experiment II: Self-supervised Deep Supervision on Fully-supervised Learning}
\label{ExperimentII}
\subsubsection{Datasets and Data Preparation}
In this experiment, ROSE \cite{ma_rose_2021} dataset, a public dataset with a total of 229 OCTA images consisted of 2 modalities, ROSE-1 and ROSE-2. ROSE-1 consisted of normal control and Alzheimer's disease patients. ROSE-2 consisted of a various kind of macular degenerative disease patients. However, the disease label unable to be provided by the original author due to ethical concerns. The train and test datasets were explicitly provided, thus the experiment were evaluated on the exactly same test dataset as the original work.

\subsubsection{Benchmarking Methods} To validate the effectiveness of the self-supervised deep supervision mechanism on the fully-supervision task, the coarse stage network of OCTA-Net were augmented by attaching adversarial attention gate into each decoder layer of both branches of the network, combined with the Discriminator network. The fine stage of the network were left as is. This 
variation of model modification is called \textbf{OCTA-Net\textsubscript{AAG-SSDS}}. The comparison was made against the original OCTA-Net results reported in the original work.

\subsubsection{Model Training and Evaluation Methods} The experimentation settings in the original work were replicated to perform a direct comparison against the reported score in the original work. The number of training epoch was fixed to 200 for both the coarse and the fine stage of the network with a batch size of 2. Adam optimizer with learning rate and weight decay parameters were set to 0.0005 and 0.0001 respectively, incorporated with polynomial learning rate scheduler with the power factor of 0.9.

\subsubsection{Performance Analysis} To show that the benefit of our method is not limited to the weakly-supervision task, the comparison was made by comparing the OCTA-Net against its deeply supervision augmented version. The results in Table \ref{table:2} show the superiority of the augmented our method for most of the cases. However, the exception were found on \textbf{ROSE-1 DVC}, where our method achieved a much lower segmentation performance compared to Ma \textit{et al}. OCTA-Net. \textcolor{black}{We speculated that the reason being the expert annotation of DVC level microvascular is limited to the area around FAZ, causing our method which relies on confidence level of the predicted segmentation map to underfit the vessel that extends further from the area.}
\begin{table}[h]
\captionsetup{width=\columnwidth}
\resizebox{\columnwidth}{!}{
\ra{1.3}\begin{tabular}{@{}lcccc@{}}
\toprule[1.2pt]
                    & \multicolumn{3}{c}{\textbf{ROSE-1}} & \textbf{ROSE-2} \\
                    \cmidrule(l){2-4}
                    \cmidrule(l){5-5}
\textbf{Method}              & \textbf{SVC}     & \textbf{SVC+DVC} & \textbf{DVC}    & \textbf{SVC}    \\
                    \cmidrule{1-1}
                    \cmidrule(l){2-4}
                    \cmidrule(l){5-5}
OCTA-Net\cite{ma_rose_2021} & 76.97  & 75.76  & \textbf{70.74} & 70.77 \\
\textbf{OCTAve\textsubscript{OCTA-Net} (Our method)}   & \textbf{78.03}  & \textbf{81.42}  & 62.55 & \textbf{71.18} \\
\bottomrule[1.2pt]
\end{tabular}}
\caption{Vessel segmentation performance (Dice’s Coefficient) in the fully-supervised learning setting on ROSE-1 and ROSE-2 datasets.}
\label{table:2}
\end{table}

\section{Discussion \& Conclusion}

\subsection{Effect of the Dynamic Weight Term Alteration on the Model Training}
\label{effect of alpha0}

\begin{figure}
    \centering
    \includegraphics[width=\columnwidth]{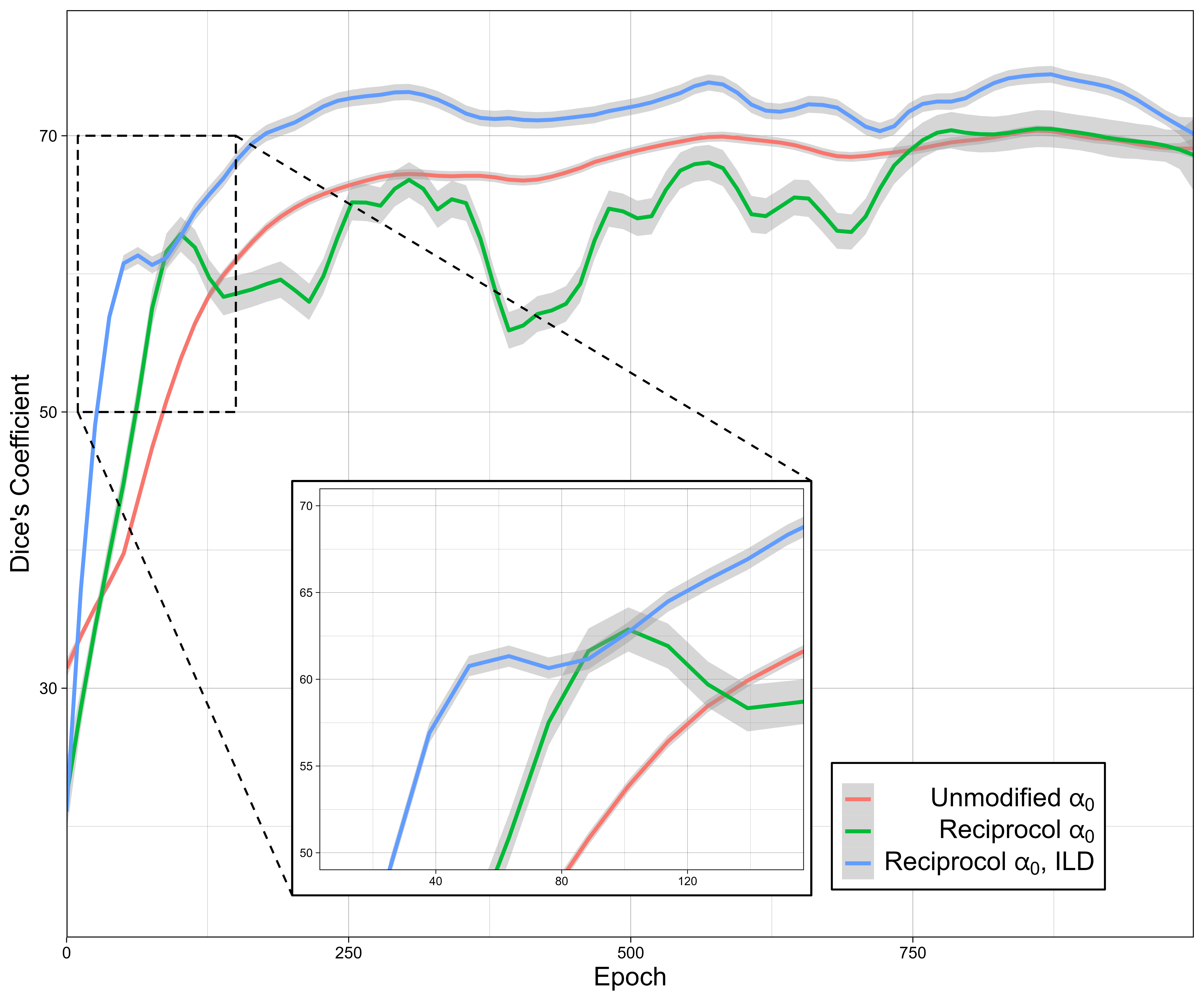}
    \caption{Dice's coefficient on validation set over the training epoch of the different $\alpha_0$ modifications on \textbf{OCTA-500 3M FULL} dataset, demonstrating the transition in the segmentation performance in the early stage of model training.}
    \label{fig:abla_alpha}
\end{figure}

For the discussion about the effectiveness of the reciprocal $\alpha_0$ over the unmodified version, we showed the segmentation performance of each method: Valvano et al. unmodified method, Reciprocal $\alpha_0$, and Reciprocal $\alpha_0$ with Inter-layer divergence loss (our method). As stated in the Section \ref{sectionIIIsubsectionA}, we traded the training stability for the faster model training, which the reciprocal version can achieve a higher segmentation performance in the early training stage as seen in the Fig. \ref{fig:abla_alpha} before destabilizing, albeit regained the stability in the late stage. However, to our anticipation, the early boost in the performance was continued by the use of a self-supervised deep supervision mechanism through inter-layer divergence loss, which can lead to the significant performance increase and the faster training compared to the previous two.

\subsection{Error Propagation Drawback from the Self-supervised Learning}
Our proposed method is a delicate system where weakly-supervised, adversarial, and self-supervised learning work in an ensemble and must be performed in harmony without disrupting the balance. Otherwise, the model may be stuck in the bad local optima. Thus, being the reason for the additional dynamic weight term (Equation \ref{kappa_term}) we proposed to regulate the self-supervised loss optimization.

While our method is superior to the original scribble-base state-of-the-art method for most cases, as observed in the results of 100\% and 50\% scribble availability variation in Table I. The experiments performed on the extreme edge case, where only 10\% of the training samples with labels were available, revealed the possible drawback of our proposed method, where self-supervision might fail and disrupt the overall performance of the model training. Hence, the amount of data annotation available needs to be taken into account for the application of our method to other tasks to avoid the performance drop due to the error propagation in the early stage.

\subsection{Conclusion}
In this work, we present a novel scribble-base weakly-supervised framework called OCTAve that utilizes adversarial deep supervision and the novel self-supervised deep supervision. It outperforms the current state-of-the-art scribble-base weakly supervised and state-of-the-art fully-supervised method on the 2D en face OCTA vessel segmentation task on the public datasets (OCTA-500 and ROSE). OCTAve is carefully designed to make three learning objectives work together in harmony, which results in a robust framework that can retain its performance advantage even if 50\% of the label is missing from the dataset. We conducted the experiment and evaluation of the segmentation performance for the weakly-supervised learning method on the OCTA-500 dataset with statistical analysis, revealing our method performance advantage that can make the lightweight UNet achieve the same segmentation performance as the heavyweight ResNeSt50-backbone UNet, while also increasing the performance of the heavyweight one. Additionally, we conducted a replica of a fully-supervised experiment on ROSE dataset to investigate the application of our proposed method other than the weakly-supervised learning and compared it against the state-of-the-art of such dataset, resulting in our method's superiority in the segmentation performance; thus, signifying the impact of our proposed method on the overall area of 2D en face OCTA vessel segmentation in both low and high data availability scenarios.


\appendices


\bibliographystyle{ieeetr}
\bibliography{references}

\end{document}